\documentclass[utf8,english]{lni}

\usepackage{fancyhdr}
\usepackage[normalem]{ulem}
\usepackage{graphicx}
\usepackage{tabularx}
\usepackage{adjustbox}
\usepackage{listings}
\usepackage{float}
\usepackage{parcolumns}
\usepackage{pdfpages}
\usepackage{xcolor}
\usepackage{xurl}
\usepackage{todonotes}
\usepackage{booktabs}
\RequirePackage[l2tabu, orthodox]{nag}
\RequirePackage{glossaries}
\RequirePackage{tcolorbox}
\usepackage{bbding}
\usepackage{pifont}
\usepackage{wasysym}
\usepackage{makecell}
\usepackage[math]{blindtext}
\usepackage{mwe}
\RequirePackage{listings}
\RequirePackage{float}
\usepackage{blindtext}
\usepackage[numbers]{natbib}

\graphicspath{ {./Figures/} }
\DeclareGraphicsExtensions{.png,.pdf}

\newacronym{MMU}{MMU}{Memory Managment Unit}
\newacronym{TLB}{TLB}{Translation Lookaside Buffer}
\newacronym{MPU}{MPU}{Memory Protection Unit}
\newacronym{PMP}{PMP}{Physical Memory Protection}
\newacronym{LUT}{LUT}{Look-Up Table}
\newacronym{IPC}{IPC}{Inter Process Communication}
\newacronym{OS}{OS}{Operating System}
\newacronym{EOS}{EOS}{Embedded Operating System}
\newacronym{RTOS}{RTOS}{Real Time Operating System}
\newacronym{DoS}{DoS}{Denial of Service}
\newacronym{RISC}{RISC}{Reduced Instruction Set Computer}
\newacronym{CHERI}{CHERI}{Capability Hardware Enhanced RISC Instructions}
\newacronym{AWS}{AWS}{Amazon Web Services}
\newacronym{LLVM}{LLVM}{Low Level Virtual Machine}
\newacronym{FPGA}{FPGA}{Field Programmable Gate Arrays}
\newacronym{TCP}{TCP}{Transmission Control Protocol}
\newacronym{IP}{IP}{Internet Protocol}
\newacronym{IoT}{IoT}{Internet of Things}
\newacronym{API}{API}{Application Programming Interface}
\newacronym{OTA}{OTA}{Over The Air}
\newacronym{ACL}{ACL}{Access Control List}
\newacronym{TCB}{TCB}{Trusted Computing Base}
\newacronym{TLS}{TLS}{Thread Local Storage}
\newacronym{LoC}{LoC}{Lines of Code}
\newacronym{MMIO}{MMIO}{Memory Mapped Input Output}
\newacronym{PFN}{PFN}{Page Frame Number}
\newacronym{ISA}{ISA}{Instruction Set Architecture}
\newacronym{FDT}{FDT}{Flattened Device Tree}

\newcommand{\note}[3]{{\color{#3}[ #1---#2 ]}}
\renewcommand{\note}[3]{}

\newcommand{\amnote}[1]{\note{#1}{Alfredo}{teal}}

\newcommand\blfootnote[1]{%
  \begingroup
  \renewcommand\thefootnote{}\footnote{#1}%
  \addtocounter{footnote}{-1}%
  \endgroup
}

\begin{document}
\title[]{Case Study: Securing MMU-less Linux Using CHERI}
\editor{Herausgeber et al.}
\booktitle{Software Engineering 2024 Workshops}
\yearofpublication{2024}

\author[Hesham Almatary \and Alfredo Mazzinghi \and Robert N. M. Watson]
{Hesham Almatary\footnote{Capabilities Limited, U.K.
\email{heshamalmatary@capabilitieslimited.co.uk}} \and
Alfredo Mazzinghi\footnote{University of Cambridge, U.K.
\email{am2419@cam.ac.uk}} \and
{Robert N. M. Watson\footnote{Capabilities Limited, U.K.
\email{robert@capabilitieslimited.co.uk}}
}}

\maketitle
\begin{abstract}

MMU-less Linux variant lacks security because it does not have protection or isolation mechanisms. It also does not use MPUs as they do not fit with its software model because of
the design drawbacks of MPUs (\ie coarse-grained protection with fixed number of protected regions).  We secure the existing MMU-less Linux version of the RISC-V port using CHERI.
CHERI is a hardware-software capability-based system that extends the ISA, toolchain, programming languages, operating systems, and applications in order to provide complete pointer and memory safety.
We believe that CHERI could provide significant security guarantees for high-end dynamic MMU-less embedded systems at lower costs, compared to MMUs and MPUs, by:
1) building the entire software stack in pure-capability CHERI C mode which provides complete spatial memory safety at the kernel and user-level,
2) isolating user programs as separate ELFs, each with its own CHERI-based capability table; this provides spatial memory safety similar to what the MMU offers (\ie user programs cannot access each other's memory),
3) isolating user programs from the kernel as the kernel has its own capability table from the users and vice versa, and
4) compartmentalising  kernel modules using CompartOS' linkage-based compartmentalisation. This offers a new security front that is not possible using the current MMU-based Linux, where vulnerable/malicious kernel modules (\eg device drivers) executing in the kernel space would not compromise or take down the entire system.
These are the four main contributions of this paper, presenting novel CHERI-based mechanisms to secure MMU-less embedded Linux.
\end{abstract}

\blfootnote{\textbf{Copyright $\copyright$ 2024 for this paper by its authors.\\
Use permitted under Creative Commons License Attribution 4.0 International (CC BY 4.0).}}
\blfootnote{\textbf{This paper is accepted at AvioSE'24.}}


\section{Introduction}
\label{contributions}

Linux is the most widely deployed \gls{OS} in the world. It is the base of Android, cloud computing, and millions of \gls{IoT} devices. The current MMU-less embedded Linux variant lacks security because either the underlying processor does not have an \gls{MMU} or it does have it but some systems intentionally do not make use of it due to determinism, power, and simplicity requirements (typically in real-time systems). There exists other protection technologies such as TrustZone and \gls{MPU}s, but they are still not widely-used in MMU-less Linux because of their coarse-grained security features~\cite{zhou2019good} that are not suitable for embedded Linux. \gls{CHERI} tries to fill this gap by providing fine-grained security at lower overhead. In this paper, we try to enhance MMU-less Linux's security by having three main goals: first, complete spatial memory safety at the pointers level, and second, spatial compartmentalisation of user programs, libraries, and kernel modules. Pointer safety protects against 70\% of software vulnerabilities~\cite{cimpanu_2019} (even in MMU-based systems), while compartmentalisation helps limit the ``future, unknown'' attack effects to the compromised compartment instead of the entire system. The third goal is source-code compatibility which ensures minimum to no changes to the C code. This is vital for high-end, MMU-less embedded systems that are already written in millions of \gls{LoC} like Linux and its userspace. We achieve these goals by building on CHERI as the sole hardware feature to secure MMU-less Linux with four different approaches:

\begin{enumerate}
  \item Building the entire software stack in CHERI C to provide complete spatial pointer safety at the kernel and user space.
  \item Isolating user programs from each other by restricting access for each ELF to what it can reference through a restricted capability table.
  \item Isolating user programs from the kernel by taking away system permissions from user capabilities and having separate capability tables for the kernel and user applications.
  \item Compartmentalising kernel modules within the kernel itself (which cannot be realised even with an MMU), by implementing the CompartOS model~\cite{UCAM-CL-TR-976, almatary2022compartos}.
\end{enumerate}

We evaluate these approaches by porting MMU-less Linux to CHERI-RISC-V~\cite{hesham2023linux} (a CHERI implementation on top of the RISC-V hardware and software ecosystem) along with userspace that contains \textit{Busybox}~\cite{hesham2023busybox}, a simple run-time linker called \textit{uldso}~\cite{hesham2023uldso}, and the \textit{uclibc-ng}~\cite{hesham2023uclibcng} C library. All of our work is open-source. We report our experiences porting MMU-less Linux and its components to CHERI and applying the CompartOS compartmentalisation model~\cite{UCAM-CL-TR-976, almatary2022compartos} to isolate user programs and kernel modules.

The porting is still a work-in-progress and is not covering the entire Linux kernel, configurations, and code execution paths. However, it is a best-effort starting point
with minimal configurations enabled that are able to run \textit{Busybox}'s shell utilities, load simple kernel modules, and run user applications, all in CHERI C (pure-capability mode, which guarantees complete spatial memory safety).
This could serve as a future reference or base for those who want to secure other (embedded or MMU-less) operating systems using CHERI.
We hope this paper gives insights to what it takes to secure complex and high-end embedded operating systems using CHERI, and the most common software subsystems that require modifications.

\section{Background}

In this section, we give a brief overview of CHERI and what security challenges it is trying to address,
mainly pointer and memory safety and software compartmentalisation. We also describe capability-based security that CHERI brings and its advantages over
traditional \gls{ACL} and MMU-based systems that exist in UNIX systems (including Linux).

\subsection{Memory Safety}
Memory-safety software vulnerabilities have arguably been the most prevalent type of bugs in the history
of computers and software in general. In their Eternal War in Memory paper~\cite{szekeres2013sok},
Szekeres et al. argue that memory corruption bugs are one of the oldest problems in computing
and still are, regardless of the efforts spent to come up with memory-safe languages and
hardware security architectures. Stack and buffer overflows~\cite{one1996smashing} have been one of
the most crucial bugs that can be exploited to affect the integrity, confidentiality,
and availability of a computer system. Memory safety is thus a vital attack vector
in both general-purpose OSes and embedded software systems that are mostly written
in memory-unsafe languages such as C/C++ for performance and fine-grained hardware control purposes.
For example, Microsoft has recently revealed~\cite{cimpanu_2019} that 70\% of security software bugs
in their systems (\eg Microsoft Windows written in C/C++) are memory-safety related, which include
spatial and temporal memory safety.
This issue does not appear to be only specific to general-purpose OSes such as Windows, but
also in the embedded software systems. Papp et al. have done a sound security analysis
in embedded systems~\cite{Papp2015} and found out that memory-safety and programming
error attacks on embedded operating systems and firmware are the most frequent and critical.

Some countermeasures against memory safety attacks in embedded systems have been proposed~\cite{parameswaran2008embedded}.
These include hardware architecture support, static and dynamic analysis tools, compiler support,
the use of memory-safe languages, and software compartmentalisation. For example,
guard pages~\cite{cowan1998stackguard} rely on the \gls{MMU} to detect buffer overflows in general-purpose
OSes. As our target OSes do not usually use the \gls{MMU}, they tend to go for lighter-weight software stack protection solutions.
For instance, stack canaries~\cite{enwiki:1066379220} can be optionally enabled by the compiler toolchain to
detect stack overflows. Similarly, some embedded operating systems such as RTEMS and FreeRTOS implement their
own stack overflow detection in software~\cite{freertosstack21, rtemsstack21}. However, such solutions
are coarse-grained (\ie only protect the stack of a task, rather than smaller buffers or global objects)
and can be bypassed~\cite{richarte2002four}.

\subsection{Software Compartmentalisation}
\textit{Software compartmentalisation}~\cite{watson2015cheri, gudka2015clean, karger1987limiting,
provos2003preventing, kilpatrick2003privman, watson2010capsicum} is a technique to split up a large
monolithic software into smaller compartments in order to reduce the attack vector and limit the
effects of a successful attack only to the compromised compartment. Unlike vulnerability mitigation
techniques, software compartmentalisation assumes that zero-day unknown vulnerabilities always
exist and acts accordingly. The measures that are taken to apply software compartmentalisation follow the
\textit{principle of least privilege}~\cite{saltzer1975protection} by only giving the very minimum
privileges to each compartment required to perform a service. This reduces privilege escalation
attacks due  to the \textit{ambient authority} problem~\cite{lampson1973note} and, thus, maintains
the \textit{integrity} and \textit{confidentiality} of the whole system.

Traditionally, OSes compartmentalise applications into threads or processes. Each could be isolated
from each other by the \gls{MMU} or \gls{MPU}. However, the recent attacks require more scalable and fine-grained
form of compartmentalisation within the same address space (\eg in a single process or kernel). This could
be linkage modules (\eg kernel modules, device drivers, or third-party libraries) or even down to each pointer or function.
Compartmentalising such smaller components within the same address space not only includes spatial and temporal
memory safety, but also fault isolation (to maintain availability).

\subsection{Capability-based Systems}

A capability, in general, is an unforgeable token to an object in the system that authorises its
holder
to access that object with a set of permissions embedded in the capability itself. Thus, a
capability
serves as both an identification (unlike \gls{ACL} systems \eg UNIX in which identities
are separate from resources in the access lists)
and an authorisation mechanism within a protected system. The capability
system itself is only responsible for the integrity of capabilities (\ie they cannot be forged)
and
serving requests to create, copy, and revoke capabilities.
The properties of a capability give it some advantages over the ACL systems. Capability-based
systems address some of the
issues like the size of the table and the requirement to have a list of resources for each domain,
which ACL systems fail to achieve.
Furthermore, capabilities inherently adopt the notion of \textit{intentionality}; an entity (such
as a process) that
has a capability to an object is only allowed to access this object with limited access
permissions to do its job. This
solves challenging issues in ACL systems such as the \textit{confinement} \cite{lampson1973note}
and \textit{confused deputy} \cite{hardy1988confused} problems.
Overall, capability-based systems give some practical solutions to security and protection issues such as,
\textit{the confused deputy, the confinement problem, scalable access control management and fine-grained access control}.

There have been significant efforts to build capability-based systems in software, hardware,
programming languages, or a combination of them; most of which are introduced in Levy's capability-based
systems book \cite{levy2014capability}.
Hydra~\cite{wulf1974hydra} was the first general-purpose object-based capability system. It was
developed at Carnegie Mellon University.
The primary motivation for Hydra was to allow operating systems research and extensibility.
Some of the design choices like putting drivers in userspace and separating policies from
mechanisms in the kernel
found their way to microkernel designs~\cite{liedtke1995micro} and are still being adopted in
modern L4 microkernels such as seL4~\cite{klein2009sel4} and Fiasco.OC~\cite{L4RuntimeEnvironment2020}.
Hydra provided a new system abstraction at the time, making everything in a computer system an
object.
EROS~\cite{shapiro1999eros} introduced a new idea of revoking capabilities by versioning objects
and capabilities pointing to them.
To revoke access to an object, the version of the object would be changed, and consequently, the
mismatched versions
(during a dereference) will trigger a fault.
Capabilities in EROS are represented as nodes. A protection domain consists of a tree of nodes of
capabilities.
Each node has a fixed size of thirty-two capabilities.
Unlike seL4 (described next), EROS tries to map pages on virtual memory faults, and if it fails, it calls a
user-level fault handler
in a capability. seL4 always redirects faults to the fault handler in userspace.
seL4 is a modern microkernel with a focus on security within embedded systems. The
authors of the 2009 seL4
paper claimed it is the first general-purpose operating system to be formally verified~\cite{klein2009sel4}. This means
that the high-level kernel specification matches the C code (and in later versions, the binary itself). With the assumption that the assembly
code and the
hardware are correct, seL4 is claimed to be bug-free and would never crash.
This comes with a few caveats and assumptions~\cite{sel4proofcaveats}. For instance, seL4 does not currently protect against timing channels, DMA-related attacks, or further exploitable hardware vulnerabilities.
It also only proves the integrity and confidentially of the system with those caveats.
Adopting microkernel principles, seL4 embraces simplicity in its design and implementation. This
enabled the trusted codebase to be small enough (less than 10K LoC) that complete formal verification of the
kernel behaviour in
every possible path is performed and reasoned about to be bug-free.
This effort narrowed the gap of having a trustworthy software system. The
downside, still, is that
the hardware is assumed to be correct, which is not usually the case. For example, the verified
seL4 code is
vulnerable to recent covert channel attacks such as Spectre~\cite{kocher2018spectre}. Furthermore,
seL4 relies on conventional \gls{MMU} to provide
isolation, which largely lacks determinism and have coarse-grained memory protection granularity of 4 KiB.
That is, seL4 will not be a good fit for embedded systems that require determinism and small size protection units
(\eg 4 bytes \gls{MMIO} registers).

\subsection{CHERI Overview}
\gls{CHERI} is a modern capability-based \gls{RISC} architecture that is being developed by the University of Cambridge and SRI International.
It is both hardware (as an \gls{ISA} extension) and software (toolchain, programming languages, operating systems, and applications).

The main principles \gls{CHERI} is designed around are:
\begin{itemize}
    \item \textit{Principle of least privilege}: which motivates the idea of reducing privilege
    rights
    to a piece of software as much as possible.
    \item \textit{Principle of intentional use}: by naming the privilege a software application
    uses, rather than giving it full privilege accesses and let it choose what privileges to use implicitly.
\end{itemize}

\gls{CHERI} is designed with these principles in mind such that a software application that runs on top
of \gls{CHERI} inherently maintains capability attributes.
This helps in reducing the access rights an attacker has and consequently minimises the attack
surface.
There are two main protection models in CHERI: pointer safety and software compartmentalisation.
Pointer safety is the primary application of CHERI in C/C++ languages. This mostly
relies on the
compiler toolchain
to map C/C++ pointers into CHERI's memory capabilities. In the current implementation of CHERI, the
toolchain is
LLVM/Clang. There are two main modes a CHERI-aware C/C++ code can be compiled in: compatible (or hybrid) and
pure-capability modes (or simply CHERI C). Hybrid mode enables users to manually select pointers to protect,
while CHERI C automatically protects all pointers.

The CHERI hardware uses a fat-pointer representation for capabilities, along with a hardware-managed tag bit that determines whether the capability is valid.
In particular, a CHERI capability encodes the \textit{base} and \textit{top} of the region in which the capability can be dereferenced.
A number of permission bits determine how the capability can be used (\eg read-only, read-write, etc.).
Tagged memory is used to maintain the tag bit for capabilities stored in memory.
The architecture uses the tag bit to enforce the CHERI capability integrity and provenance validity properties.
The \gls{ISA} maintains the capability monotonicity property, whereby instructions can only narrow the permissions and bounds of any given capability.
CHERI hardware capabilities form the primitive upon which it is possible to implement language-level spatial memory safety and software compartmentalisation models.

Apart from pointer safety, CHERI provides the flexibility for software developers to define their own representation
of a software compartment in order to logically split large monolithic software systems into smaller
compartments. For example, CHERI could be used to compartmentalise processes, linkage-modules, static
or shared libraries, OSes and applications, etc.

\begin{table}[]
\centering
\begin{tabular}{@{}lll@{}}
\toprule
                                     & \textbf{MMU} & \textbf{CHERI/CompartOS} \\ \midrule
\textbf{Pointer safety}              & \ding{55}               & \CheckmarkBold \\
\textbf{Compartmentalisation}        & \CheckmarkBold & \CheckmarkBold \\
\textbf{Virtualisation}              & \CheckmarkBold & \ding{55}    \\
\textbf{Capability-based protection} & \ding{55}      & \CheckmarkBold \\
\textbf{In-address-space isolation} & \ding{55}      & \CheckmarkBold \\
\textbf{Protection granularity}      & 4 KiB         & 1 Byte         \\
\textbf{Isolated memory resources} & \thead{User processes and kernel}      & \thead{User processes, kernel, \\libraries, kernel modules, and pointers.} \\  \bottomrule
\end{tabular}
\caption{MMU vs CHERI comparison.}
\label{mmu_vs_cheri}
\end{table}

CHERI is fundamentally different than MMU. MMUs are implemented in hardware and managed by the OS to provide both
protection and virtualisation at 4 KiB granularity. CHERI, on the other hand, is an ISA extension that is implemented
also in hardware, but could be managed by the OS, programming languages, linkers and loader, and applications in order to provide
fine-grained pointer safety and memory protection (\eg 1 byte) and scalable software compartmentalisation (\eg isolating processes, libraries, or modules).
The main differences between CHERI and MMUs are shown in Table~\ref{mmu_vs_cheri}.
In this paper, we deploy two main applications of CHERI: 1) fine-grained memory safety in C/C++, and 2) software compartmentalisation.

CHERI is capable of replacing the MMU in terms of protection and isolation between processes. However, this is only one contribution of this paper (contribution number 2 in Section~\ref{contributions}). CHERI and CompartOS offer
more security features than MMUs. MMUs cannot provide pointer and memory safety at the programming languages level as CHERI does because of their coarse-grained page size granularity.
Furthermore, CHERI and CompartOS allow in-address-space protection; two different pointers, user applications, or kernel modules can be isolated from each other. Arguably speaking, this
cannot be done via MMUs unless the entire design and implementation of the Linux kernel and applications are going to be changed. As shown later, it takes minimal effort to do so
using CHERI/CompartOS in terms of \gls{LoC} changes, embracing source-code compatibility.
All in all, CHERI/CompartOS provide more security, scalability, compatibility in embedded systems against recent attacks even compared to MMUs.

\subsection{MMU-less Linux}
MMU-less Linux is a variant of the mainstream Linux targeting embedded applications. It is a configuration
option to enable building and running smaller and customised Linux subsystems.
Systems choose not to use the MMU for two main reasons: 1) either the underlying hardware does not have
an MMU unit, or 2) there exists an MMU, but managing it adds extra complexity, size, or does not meet
some requirements such as real-time, determinism, power consumption, etc.
MMU-less Linux could have coarse-grained privilege-separation between the kernel and user. However, there is
no spatial memory protection among user programs, kernel modules, and the kernel.
MMU-less Linux  uses light-weight binary formats for ELFs. Historically, FLAT-ELF was mostly
used but comes with restrictions such as limited number of shared libraries, and no support for dynamic loading
via \textit{dlopen}. ELF-FDPIC format tries to overcome such limitations and gets all of the usual ELF features,
but the code has to be all position independent (PIC). This allows different load segments to be independently located
in memory while still being able to share the text segment, but not necessarily data segments.
In this paper, we use RISC-V as a base for MMU-less Linux, which does not use MMU or MPU/PMP. We further
choose ELF-FDPIC which works best for our CHERI-based compartmentalisation in embedded systems.

\subsection{Potential Applications in Avionics}
While CHERI and CompartOS can be applied to mainstream baremetal and tiny embedded systems (at a cost), it
aims to secure the high-end range of \textit{mainstream} embedded operating systems (including \gls{RTOS}es), such as Linux. By mainstream,
we mean traditional embedded (operating) systems that are mostly unprotected and shy away from using MPUs and \gls{MMU}s
as they do not meet their requirements. This is in contrast to using \textit{new} security architectures that target small
systems (\eg TockOS~\cite{levy2017multiprogramming} and ACES~\cite{clements2018aces}) and require existing applications to be (re)written on top of them.
Traditional embedded operating systems are facing security concerns and need some form of protection while continuing to meet
their real-time and safety-critical requirements such as partitioning, bounded processing, high
determinism and high throughput. Developers either try to use the MPU in MCUs that
provide it, but that is not scalable or fine-grained enough for complex applications, or they go
for general-purpose processors (\eg A-class Arm  processors) that have \gls{MMU}s (that are not frequently used
due to complexity and performance non-determinism) and rather use the MPU. The area
those systems fit in is safety-critical avionics and automotive.
For example, FreeRTOS or RTEMS run on Raspberry Pi~\cite{enwiki:1072126107} or Beagle~\cite{enwiki:1053954096}
boards (embedded with Arm A-class processors) without an \gls{MMU}-based process
or protection model.
CompartOS mainly targets the low-end A-class category. Example deployed systems are:
\begin{itemize}
    \item Amazon's FreeRTOS with WiFi, TCP/IP, and Bluetooth stacks running on high-end M-class
    processors (\gls{AWS} Reference Integrations~\cite{freertosboards21aws}) and used for feature-rich \gls{IoT}.
    \item Primus Epic Avionics: Deos and ARINC 653, running on x86 and Arm's A-class~\cite{deos21ddci}.
    \item VxWorks CERT EDITION, running on NXP QorIQ~\cite{enwiki:1057508237} for automotive and avionics~\cite{vxworks21wr}.
    \item RTEMS used in NASA's Magnetosphere Multiscale (MMS) Mission~\cite{rtemsnasa22, rtemsnasa222} running on the Coldfire CPU~\cite{enwiki:1031125403}.
\end{itemize}

On the application use cases, some safety-critical standards such as ARINC 653 could greatly fit with the CompartOS model.
Most ARINC 653 implementations, however, are proprietary and closed source. While we mainly apply CHERI and CompartOS
models to Linux in this paper, we believe that they have been also applied to other avionics systems such as RTEMS
and FreeRTOS, and could easily be applied to VxWorks and ARINC 653 implementations to provide partitioning, bound processing,
and availability.

\section{Design}

\begin{figure}
    \centering
    \includegraphics[scale=1]{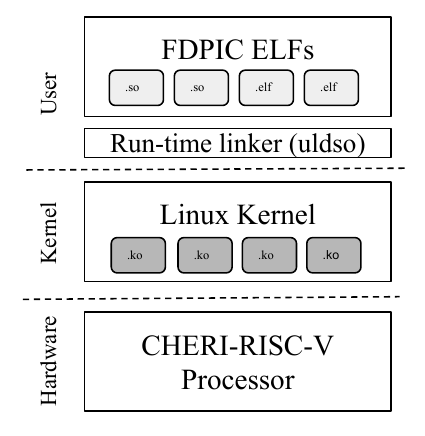}
    \caption{
     Secured MMU-less Linux using CHERI. All build and run in CHERI C. Grey boxes represent different types of compartments such
     as kernel modules (.ko) and ELF-FDPIC programs (.elf and .so).
    }
    \label{fig:elinux_runtime}
\end{figure}
\amnote{Can this be smaller? Not sure if it warrants a whole page.}

\begin{figure}
    \centering
    \includegraphics[scale=.5]{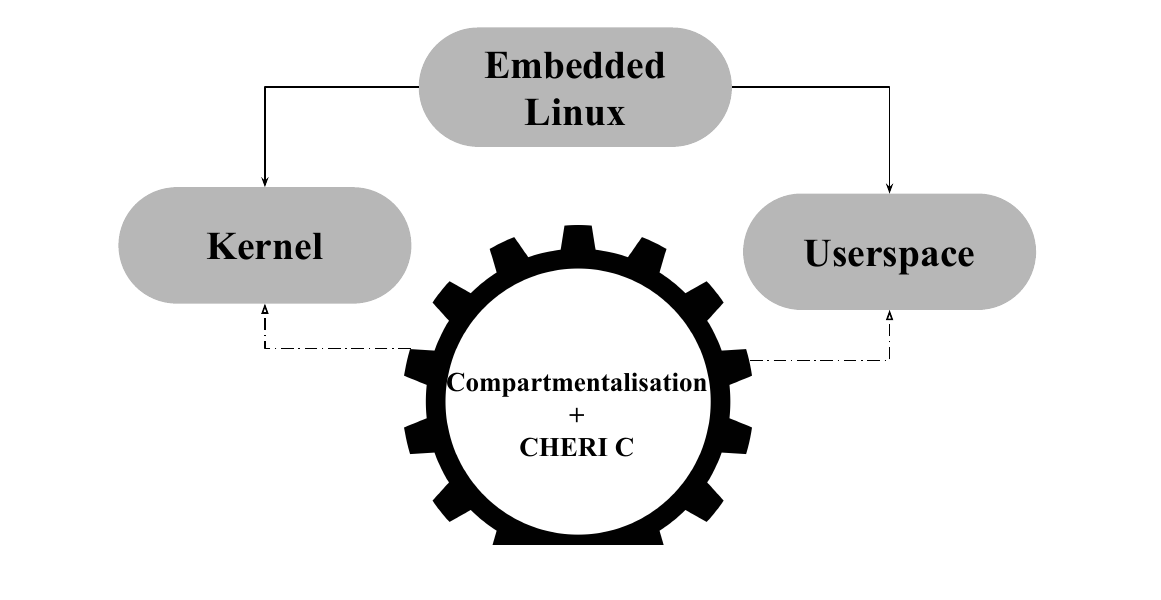}
    \caption{
     The process of securing MMU-less Linux using CHERI is divided up into CHERI C and compartmentalisation stages for each software component.
    }
    \label{fig:elinux_process}
\end{figure}

The main motivation for this work is to take an unsecure monolithic MMU-less Linux and secure it using CHERI as in Figure~\ref{fig:elinux_runtime}. In an MMU-less Linux, there is only privilege-level separation between the kernel
and userspace. There is no memory safety or separation within or among user applications or kernel modules. Malicious user applications or modules can spy on (no confidentiality) or affect the functionality of one another (no integrity),
or take down the entire system (no availability). CHERI-based memory safety and compartmentalisation provide some form of confidentiality, integrity, and availability.

The process of doing that is divided up into steps shown in Figure~\ref{fig:elinux_process}. Securing MMU-less Linux is done in two steps: \textit{CHERIfying} the code and enforcing a compartmentalisation model. \textit{CHERIfying} means that we port the existing C code into CHERI C (or pure-capability mode) where every pointer is a capability. The source-code has to be available to re-compile. This provides complete spatial memory safety within the entire software stack. It helps with the confidentiality and integrity of the software system. However, it does not necessary help with availability; a CHERI security violation in the kernel may still bring down the whole system. That is where compartmentalisation is important. Compartmentalisation divides up the entire monolithic kernel (and userspace) into smaller compartments; each compartment is contained. A fault in one compartment does not affect the rest of the system. This is crucial for third-party (and likely untrusted) user libraries and kernel modules such as device drivers.

For CHERI C, each software subsystems in the kernel and userspace needs modifications in the following areas:
\begin{enumerate}
    \item \textbf{Build system:} necessary changes to provide the proper toolchain flags to enforce CHERI C memory safety.
    \item \textbf{Low-level}: mostly architecture-dependent code such as booting, handling traps, atomics, etc.
    \item \textbf{Pointers}: some/most C/C++ projects mix using integer types (\eg \textit{int} and \textit{long}) to store both integers and pointers and do computations and arithmetic on them, as this is allowed by C. This could lead to out-of-bounds memory safety bugs. CHERI C is strict and it differentiates between normal integers and address pointers as CHERI capabilities.
    \item \textbf{Provenance}: every capability has to be derived from another valid capability with the same or less bounds and permissions (monotonicity attribute). There are subsystems in the \gls{OS} that need to manually create capabilities. For instance, the boot code creates capabilities for global objects and functions from \textbf{\textit{DDC}} (root data capability) and \textbf{\textit{PCC}} (root code capability), respectively. Similarly, new capabilities for \gls{MMIO} need to be created for device drivers.
    \item \textbf{Allocators}: dynamic memory allocators (\eg \textit{malloc()}) need to be modified to return valid bounded capabilities with the correct permissions. Extra changes could also be applied to ensure temporal memory safety (\eg revocation, memory zeroing, etc).
\end{enumerate}

For compartmentalisation, the following subsystems need to be changed:
\begin{enumerate}
    \item \textbf{Build system}: necessary changes to provide the proper toolchain flags to enforce CompartOS compartmentalisation for compartments.
    \item \textbf{Start-up}: to be able to build a capability table for each compartment representing its protection domain and interface with other compartments.
    \item \textbf{Cross-compartment calls}: to create capabilities that perform inter-compartment calls, grant them to callers, and emit necessary trampolines to perform compartment and protection domain switches.
    \item \textbf{Fault handling and recovery}: an implemented custom policy to handle CHERI C security violations per compartment.
\end{enumerate}

In the following section, we discuss the implementation details of the specific MMU-less Linux software subsystems we worked with.

\section{Implementation}

We have used the upstream Linux (version 6.1) and the RISC-V port without MMU as a baseline. Userspace consists of \textit{Busybox}~\cite{busybox2023} and a simple run-time linker~\cite{uldso2023}, along with \textit{uclibc-ng}~\cite{uclibcng2023} as a user C library. We run the complete
software stack on \textit{QEMU}. This gets us a shell and all \textit{Busybox} utilities. Table~\ref{table:purecap_loc_changes} demonstrates the modified systems and the \gls{LoC} changes for each.

\begin{table}[t]
\centering
    \begin{adjustbox}{max width=\columnwidth}
        \begin{tabular}{l|cccc}
            \toprule
            \multicolumn{1}{r}{}
            & \multicolumn{1}{c|}{Linux Kernel}
            & \multicolumn{1}{c|}{uldso}
            & \multicolumn{1}{c|}{uclibc-ng}
            & \multicolumn{1}{c|}{Busybox}
            \\ \midrule
   \textbf{Build system} & 3/3 & 1 & 0 & 6 \\
\textbf{Low-level} & 1926/893 & 16 & 309/67 & 5\\
\textbf{Pointers} & 324/280 & 0 & 0 & 0\\
\textbf{Syscalls} & 553/543 & 0 & 37/37 & 0\\
\textbf{Provenance} & 26/13 & 0 & 0 & 0\\
\textbf{Misc} & 46/36 & 0 & 219 & 0\\
\textbf{Compartmentalisation} & 468/28 & 80 & 0 & 0\\
\textbf{Allocators} & 0 & 0 & 18 & 0\\
        \bottomrule
        \end{tabular}
    \end{adjustbox}
    \caption{Number of inserted/deleted lines to secure embedded MMU-less Linux using CHERI.}
\label{table:purecap_loc_changes}
\end{table}

\begin{figure}
    \includegraphics[scale=0.3, width=\columnwidth]{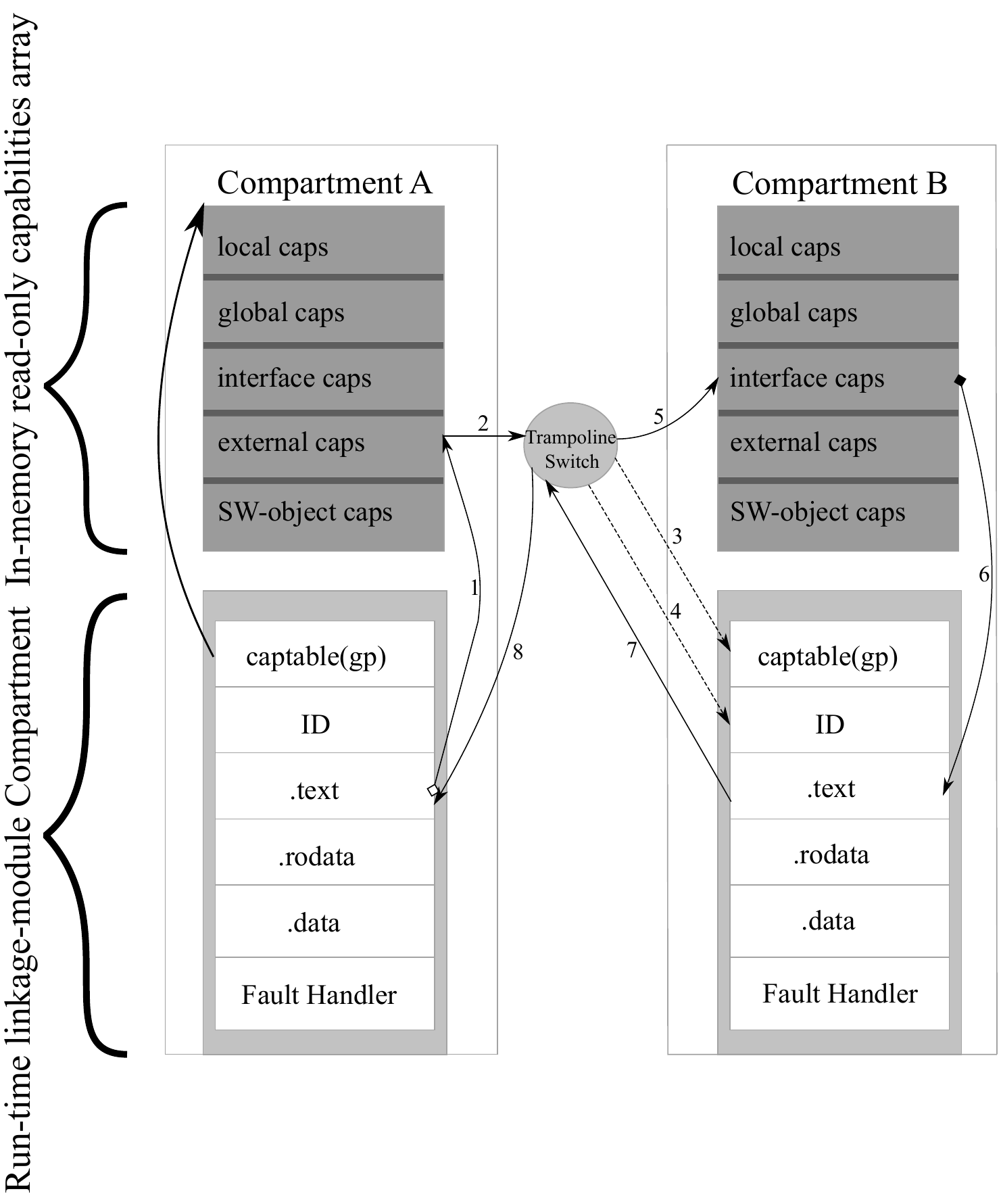}
    \caption{A runtime example of a compartmentalised Linux kernel system and a protection domain switch.
       Each compartment is this case is a kernel module. Filled lines
        are capability references, while dashed lines are operations. Numbered edges denote the sequence
        to perform an external
        call to another compartment, which triggers a domain switch. The switch starts with a
        function call that references
        a capability from the externals capability list (\#1). If the external capability
        is present and valid, it points to a small, read-only trampoline (\#2) that performs
        compartment switches by setting the new captable
        and compartment ID (\#3 and \#4) after storing the caller's context. It then jumps via the
        interface capability (\#5) provided by the target compartment.
        This interface capability points to the function within its associated compartment (\#6).
        Upon its return (\#7),
        the trampoline restores the caller's context, captable and ID, then returns back to
        the caller function (\#8).
    }
    \label{fig:comp_illustration}
\end{figure}

\begin{figure*}[ht]
   \centering
    \includegraphics[scale=.7]{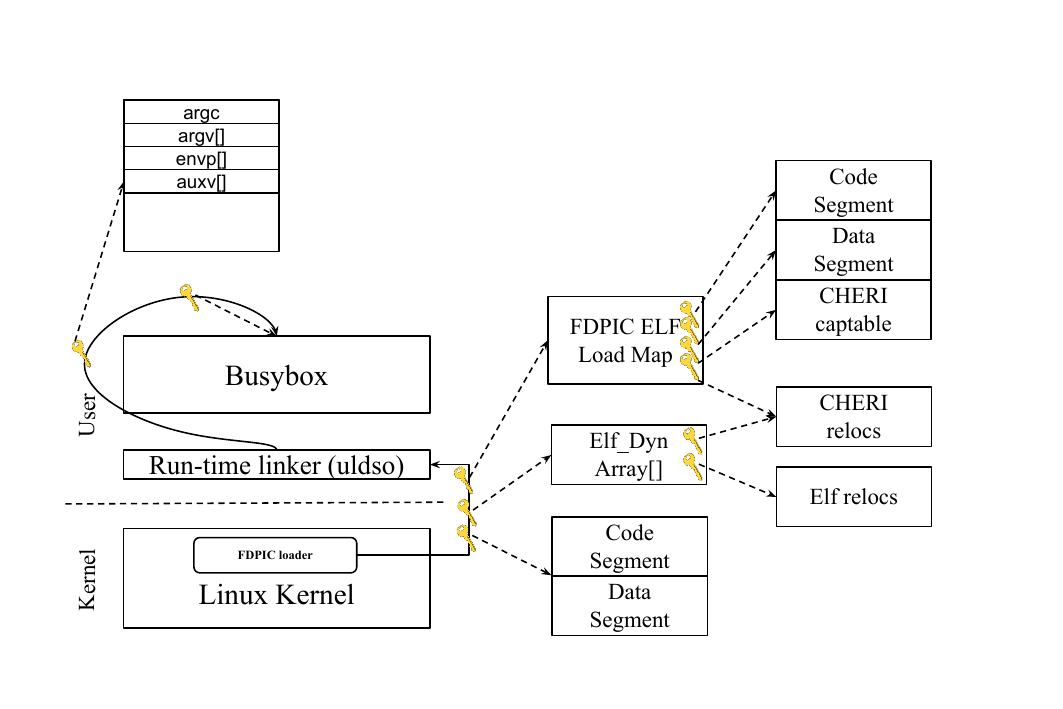}
    \caption{A runtime example of a compartmentalised Busybox userspace.
       The keys are capabilities. Solid lines/curves represent program transfer flow.
       Capabilities on these lines are passed as register arguments.
       Dashed lines show what object each capability holds an authority to.
    }
    \label{fig:comp_user}
\end{figure*}

\begin{lstlisting}[float,floatplacement=H,basicstyle=\small,caption={Compartment trampoline pseudo
        code},label={lst:tramp_code},frame=single,captionpos=b]
    xPortCompartmentEnterTrampoline:
    % Metadata
    .Lfunc:      .zero CAP_SIZE % Callee's function capability
    .Lcaptable:  .zero CAP_SIZE % Callee's capability table
    .Lcompid:    .zero CAP_SIZE % Callee's compartment ID
    // Compartment/Domain Switch
    Save caller's context (register set, GP, compid)
    Setup new GP/captable
    Set currentCompartmentID = Lcompid
    (Optional) Restrict/bound the callee's stack
    Call the destination function
    Restore caller's context (register set, GP, compid)
    Return to the caller compartment
\end{lstlisting}

\subsection{Kernel}
  \textbf{Low-level}: The boot code which is usually written in assembly is the first subsystem that needs to be modified. First, \textbf{\textit{PCC}} and \textbf{\textit{DDC}} permissions need to be restricted (\ie not to include execute and/or load/store permissions).
  Furthermore, the instructions and registers need to be modified to deal with capability registers rather than integer registers. This includes atomic instructions as well.
  The code starts in hybrid mode, then the capability table is written for all the global pointers in the Linux kernel. These are driven from \textit{\textbf{PCC}} (for code capabilities like functions) and \textit{\textbf{DDC}} (for data capabilities like variables,
  arrays, and any other data objects). The exception table and trap register also need to be set up to be capability-aware. Finally, the kernel jumps to the C code in pure-capability mode.
  Exception code is also written in assembly, but only instructions and registers need to be modified to deal with capabilities. There are other higher-level changes in the architecture dependent code, still. For example,
  the types and sizes of the registers, pointer variables, stack pointer, program counter, return address, frame pointers, etc., all needed to be capabilities instead of integers. Similar changes in the memory management code that converts \gls{PFN} to
  virtual/physical addresses also needed to be return capabilities.

  \textbf{System calls}: Generic Linux system calls macros treated all arguments and return values as type \textit{long}. This needed to be changed to be pointers or capabilities. There are some specific system calls (used by \textit{Busybox}) that also needed to be changed.
  For instance, \textit{ioctl} system calls and their kernel functions receive an argument from the user that could hold pointers. Even though we changed the lowest level of the system call macros to receive capabilities,
  further \textit{ioctl} handler functions still had the argument type as \textit{long}. This was a bit disruptive to change the functions prototypes and implementations from \textit{long} to pointers (\textit{uintptr\_t}, which could hold a pointer or an integer). Similarly, \textit{prctl} and \textit{clone}
  system calls.

  Another important system call that needed to be changed is \textit{mmap}. For MMU-less Linux, it is mainly being used to allocate memory; it does not use any virtual mappings as there is no paging. Unlike the previous system calls, \textit{mmap} returns a pointer to an allocated memory. The return type needed to also be changed from \textit{long} to a pointer type that could hold a bounded valid capability (if memory allocation is successful). We also mapped Linux's VM permissions passed to \textit{mmap}
  (\eg RWX) to CHERI permissions and disabled access to system registers for returned \textit{mmap} capabilities. Such permission flags passed to \textit{mmap} are ignored in MMU-less Linux.

  \textbf{Pointers}: This is architecture independent code that is confusing integers with pointers and addresses. The code usually uses \textit{long} as a type (or casts to it) instead of using proper types such as pointers (\eg \textit{char *} or \textit{intptr\_t})
  when dealing with addresses and pointers (\eg defining base and end addresses or performing pointer arithmetic) that get de-referenced later. This causes capabilities to lose their metadata (including tags), and thus, trigger security violations when accessed later.
  Example code is the memory management subsystem that allocates and frees pages and defines memory regions and blocks (with base addresses and sizes). Another example is data structures such as radix-tree, rbtree, and maple-tree where they perform
  casting operations to retrieve parents from nodes, or convert nodes to entries (and vice versa) and perform some masking on pointers. File handling code also needed to be changed a bit. Functions that take file descriptors as arguments
  and return a buffer or struct addresses needed to return capabilities instead of \textit{long} integers.

  \textbf{Provenance}: First, drivers that deal with \gls{MMIO} devices need to create capabilities for each region. For MMU-less Linux, this was in the FDT code where it gets the base addresses of devices and their \gls{MMIO} sizes from the \gls{FDT} nodes. We then
  create CHERI capabilities for those which are returned and held in their drivers as pointers. Similarly, \textit{ioremap} needed to be changed to return new valid capabilities. Second, we needed to create capabilities on the fly for some
  ELF loading code. The current ELF specification uses \textit{long} types to hold addresses/pointers (\eg \textit{Elf64\_Phdr.p\_vaddr} that holds addresses of program headers and segments). We do not want to change the ELF specification itself in this
  occasion, so we had to create data capabilities for the code that is offsetting ELF segments while loading ELF binaries.

  \textbf{Misc}: This includes some changes that are not part of the above categories. For instance, Linux sometimes uses reserved unused variables for padding within some structs. The number (or types) of such unused variables needed to change to meet
  the required alignment, offsets, or padding. We also needed to use our own versions of string functions such as \textit{memcpy}, \textit{memmove}, \textit{memcmp} to preserve tags. Other string-related functions such as \textit{strscpy} and \textit{strncpy} (from user), and \textit{strlen} did out-of-bounds reads as well for lengths shorter than the size of \textit{long}, so this needed to be fixed to read byte-by-byte.

  \textbf{Compartmentalisation in the kernel} is done on loadable modules. Each kernel module could be built as a compartment by providing a new compiler flag in order to reference all function and variable references GP-relative, instead of PC-relative, as
  discussed in the CompartOS paper~\cite{almatary2022compartos}. Furthermore, each module has its own capability table, separate from the kernel's. External symbols referenced within the modules that belong to the kernel or other external modules are created and added
  to the module's capability table in its external capability section (see Figure~\ref{fig:comp_illustration} and Listing~\ref{lst:tramp_code}). References to external symbols trigger a protection domain switch as shown in Figure~\ref{fig:comp_illustration}. The \gls{LoC} changes to support the CompartOS compartmentalisation model in the kernel are shown in Table~\ref{table:purecap_loc_changes}. This also includes handling new CompartOS GP-relative ELF relocations.

  \textbf{Compartmentalisation in userspace} is realised by the fact that each CHERI C ELF-FDPIC program  has its own separate capability table; communication between different ELFs and processes is left to Linux IPC mechanisms, and is not linkage-based
  external capabilities (as in CompartOS)\amnote{This sentece reads weirdly}. The compartmentalised userspace is shown as in Figure~\ref{fig:comp_user}. The kernel's ELF-FDPIC loader sets up the process environment and memory. It allocates a single chunk of memory
  for each ELF image, including \textit{uldso} and \textit{Busybox}. It then transfers the program flow to userspace along with restricted number of capabilities passed as arguments. For \textit{uldso}, the ELF-FDPIC loader hands over capabilities to the dynamic ELF section in order to perform
  the required ELF relocations. This includes relocations to normal ELF symbols and CHERI capabilities. It also gives \textit{uldso} a capability to the ELF-FDPIC's load map structure, which itself includes subset capabilities to \textit{Busybox's} loaded ELF segments.
  Once the relocation process is completed by \textit{uldso}, the program flow transfers to \textit{Busybox} with another set of restricted capabilities. The first capability is the root capability that \textit{only} covers the entire \textit{Busybox's} ELF loaded memory segments.
  The root capability gets used later (at process startup) to derive fine-grained capabilities for all global variables and functions and populates a confined capability table for \textit{Busybox} and \textit{uclibc-ng}; similar to the Linux's boot code. The second capability is the stack pointer which points to the process'
  arguments, ELF \textit{auxv}, \textit{envp}, etc., as expected by the POSIX environment. This stack capability is set up by the ELF-FDPIC loader and gets passed to \textit{uldso} which hands it over to \textit{Busybox} as it is. Thus, each user process is only restricted by these two capabilities and cannot access anything else (\eg other processes' memories).

\subsection{Userspace}
For the userspace, we use \textit{Busybox} which is widely used in embedded MMU-less Linux to provide some embedded UNIX/POSIX utilities. We also use \textit{uclibc-ng} as a C library for simplicity, which is also widely used in MMU-less Linux. Finally, a simple run-time linker was used called \textit{uldso}.

\subsubsection{uldso}

\textit{uldso} is a very simple and light-weight dynamic linker. It relies on Linux's ELF-FDPIC loader which is used for MMU-less systems. \textit{uldso} assumes that the code is compiled position-independent with the \textit{PIE} flag and the ELF-FDPIC loader has set up some registers and process segments at process startup. It then uses this information to fix up dynamic relocations.
CHERIfying \textit{uldso} enables both CHERI C user applications, and ELF-based compartmentalisation (user programs and libraries). The CHERI-LLVM toolchain emits a capability table and a relocation ELF section per ELF. Thus, this effectively sandboxes or
compartmentalises each loaded ELF, and isolates it from one another; which gets us spatial memory isolation among ELFs, similar to what the MMU provides. Furthermore, supporting CHERI C ELFs gets us in-address-space spatial memory safety that MMU-based
processes cannot have.
The code changes to support CHERI are quite minimal. Startup RISC-V assembly code needed to deal with capability registers instead of integer registers, before jumping to the C's linker code. The linker simply needed to be taught about a new
CHERI relocation section and its entries, and it then relocates each relocation entry with the run-time load address of the symbol's loaded segment in memory. This relocation section gets used later when the actual ELF bootstraps during the CHERI C
initialisation process, which basically populates the capability table with valid capabilities.

\subsubsection{uclibc-ng}

The low-level part of this is quite similar to the Linux kernel. The capability table is populated at start-up, using the capability relocation info, fixed up by \textit{uldso}. Atomics, context switching, \textit{setjump/longjump}, also needed to deal with capability registers.
There are also some low-level system calls parts in RISC-V assembly (\eg \textit{vfork} and \textit{clone}) that needed to be modified to pass and return CHERI capability registers.
The system call changes were minimal, and basically required to change the assumption and types of registers, arguments, and return values from \textit{integers} and \textit{longs} to capability registers (see Listing~\ref{lst:syscall_cheri}).
Minor changes to \textit{malloc} were also required to enforce pointer-sized alignments and granularity. Furthermore, internal \textit{malloc} code was using integer types for blocks and their arithmetic, that needed to be changed to capabilities as well.

The part that required most additions is libc's string functions. This includes \textit{memcpy} family, and \textit{str*} functions. They were copied from Linux. \textit{memcpy} and \textit{memmove} needed to perform capability-aware operations to preserve tags. \textit{str*} functions
were mostly doing out-of-bounds accesses, trying to aggressively optimise performance by reading word-sized chunks of memory. We used byte-based versions of those that are less optimised but do not perform out-of-bounds accesses.

\begin{lstlisting}[float,floatplacement=H,basicstyle=\small,caption={Git diff of the uclibc-ng common system call function},label={lst:syscall_cheri},frame=single,captionpos=b]
-long syscall(long sysnum, ...)
+uintptr_t syscall(long sysnum, ...)
 {
-       unsigned long arg1, arg2, arg3, arg4, arg5, arg6;
+       uintptr_t arg1, arg2, arg3, arg4, arg5, arg6;
        va_list arg;
        va_start (arg, sysnum);
-       arg1 = va_arg (arg, unsigned long);
-       arg2 = va_arg (arg, unsigned long);
-       arg3 = va_arg (arg, unsigned long);
-       arg4 = va_arg (arg, unsigned long);
-       arg5 = va_arg (arg, unsigned long);
-       arg6 = va_arg (arg, unsigned long);
+       arg1 = va_arg (arg, uintptr_t);
+       arg2 = va_arg (arg, uintptr_t);
+       arg3 = va_arg (arg, uintptr_t);
+       arg4 = va_arg (arg, uintptr_t);
+       arg5 = va_arg (arg, uintptr_t);
+       arg6 = va_arg (arg, uintptr_t);
\end{lstlisting}

\subsection{Busybox}
\textit{Busybox} was quite cleanly written, meaning that almost no changes were required at all to build and run it in CHERI C mode. Minor changes to low-level assembly code were required to read capability variables instead of integers.
Most importantly, we can use \textit{insmod} command and friends to load kernel modules as CompartOS compartments.

\section{Future Work}
This work is still in progress and has been only evaluated on \textit{QEMU} for functional correctness and applicability. We plan to keep enhancing it by cleaning up the code and enabling more Linux features (\eg networking, VirtIO, devices, etc).
We will also aim to run on actual CHERI-RISC-V and/or Morello hardware in order to perform performance evaluation and run stock benchmarks, similar to \textit{CheriFreeRTOS}~\cite{almatary2022compartos}.
On the security side, we will aim to reproduce some of the Linux memory safety CVEs and see if CHERI could protect against them, in terms of integrity, confidentiality, and availability. As Linux in general is a demanding high-end
system, evaluating scalability in terms of the number of user programs, libraries, and kernel modules would also be on our future work.
We have not attempted to use CHERI sub-object bounds enforcement. This is an alternative compilation mode that automatically narrows capability
bounds for pointers to individual structure members and arrays, restricting the possibility of intra-object memory corruption.
Consider, for example, the C structure in Figure~\ref{fig:subobject-example}. In a pure-capability program, a pointer to \texttt{struct sensitive\_data} produced by an allocator is mapped to a capability bounded to the size of the object (\ie the \textit{length} of the capability is \texttt{sizeof(struct sensitive\_data)}). In the default CHERI compilation mode, pointers to sub-object members inherit the bounds of the object capability. This means that the capability for the \texttt{buffer} field has the same bounds as the parent structure. As a result, a buffer overflow of the \texttt{buffer} array is not detected as long as it remains within the bounds of the parent structure. This is clearly a limitation as intra-object memory safety is not provided.

\begin{figure}
  \centering
  \begin{lstlisting}[language=C]
  struct sensitive_data {
    int some_value;
    char buffer[128];
    struct sensitive_data *next;
  };

  struct sentitive_data *p;
  // Does not trigger a fault without sub-object bounds
  p->buffer[130] = 'a';
  // Triggers a fault with and without sub-object bounds
  p->buffer[200] = 'a';
  \end{lstlisting}
  \caption{Example of a structure that benefits from CHERI sub-object bounds enforcement. Note that, without sub-object bounds enforcement, buffer overflows (and underflows) via the \texttt{p->buffer} pointer are not prevented as long as they remain within the bounds of \texttt{struct sensitive\_data}. }
  \label{fig:subobject-example}
\end{figure}

The sub-object bounds compilation mode provides superior security properties, however it exhibits larger friction for porting existing code bases \cite{UCAM-CL-TR-949} due to some problematic C-language idioms. It should be noted that the CheriBSD pure-capability kernel already provides sub-object bounds enforcement, therefore there is reason to believe that the same level of support can be achieved in Linux. Previous experience with the CheriBSD pure-capability kernel has found that the majority of the portability issues arise from the use of \texttt{containerof()} patterns, where a pointer to the container structure is constructed from a pointer to a member. These kinds of patterns break in the presence of sub-object bounds, however CheriBSD has shown that it is feasible to annotate these problematic cases to selectively opt-out of sub-object bounds while maintaining the security benefits for most of the code-base.
Future work will evaluate the use of sub-object bounds to improve upon the base CHERI spatial memory safety guarantees.

Finally, we have not tried to support or evaluate temporal safety yet. Future work will significantly enhance security by applying some of the CHERI-based temporal safety techniques such as CherIvoke~\cite{xia2019cherivoke} or CapRevoke~\cite{filardo2020cornucopia}.

\section{Related Work}
ACES~\cite{clements2018aces} is an MPU-based software compartmentalisation approach that
aims to automatically create compartments at build time by statically analysing the source
code and an input security policy. It relies on off-the-shelf MPUs to enforce memory protection and instruments
the final binary with MPU-based compartment switches. However, ACES only targets simple static systems, and would not
scale  to dynamic and high-end embedded systems like embedded Linux.

MINION~\cite{Kim2018} is an MPU-based software security architecture for embedded systems that
provides memory isolation
between processes while optimising performance by avoiding performing frequent systems calls that are
often associated with multi-privilege rings and MPUs. The paper argues that frequent system calls
usually violate real-time system constraints and responsiveness. It shares the same limitations as MPU-based
techniques as with ACES.

uVisor~\cite{uvisorgithub22} also uses Arm's MPUs and is similar to MINION as it can run an RTOS and other compartments in an unprivileged ring.
However, it is different from MINION as it dynamically loads and creates compartments at runtime
and is able to place different software entities in a compartment such as interrupt handlers
or linkage-based modules, besides threads. Thus, it is not only a task-based compartmentalisation approach.

\gls{CHERI} has been under research in UNIX-based environments with \gls{MMU}, prototyped
in the CheriBSD OS (a CHERI-enabled fork of the FreeBSD OS).
CheriABI~\cite{davis2019cheriabi} is an application-level software compartmentalisation technique in CheriBSD.
The main software application in CheriABI is C/C++ language pointer
safety at the user level with a few modifications to the FreeBSD kernel.
Two compilation modes are supported for \gls{CHERI}: hybrid and pure-capability modes. In hybrid mode,
pointers are integers as usual, and only those annotated with \textit{\_\_capability} keywords are
protected by
CHERI. CheriABI falls in the pure-capability category where user processes are compiled to have all
pointers, system call arguments and allocated C objects (such as malloc and TLS) represented as
CHERI capabilities. This significantly enhances spatial memory safety in UNIX while it is still
being compatible with native UNIX processes that are not aware of CHERI.
There is still ongoing research to
have the FreeBSD kernel itself making full use of CHERI to compartmentalise the kernel components
and enforce pointer safety. This is known as a pure-capability CheriBSD kernel.
The CheriBSD kernel makes extensive use of CHERI memory safety features, including spatial and referential memory safety,
as well as sub-object bounds. In particular, new abstractions for virtual memory management are necessary to ensure
the representability of mappings. Similarly, kernel allocators enforce representable bounds.
The use of sub-object bounds shows promising results to protect from intra-object memory corruption.

CheriRTOS~\cite{xiaCheriRTOSCapabilityModel2018} is an early exploration of CHERI in small embedded
systems. Unlike CheriABI (that targets UNIX-based systems), CheriRTOS targets microcontrollers
to offer hardware protection using 64-bit compressed CHERI-MIPS capabilities. The paper shows how fine-grained memory
protection, task isolation, secure heap management, and secure cross-domain transition can be
implemented on CHERI. The authors also argue that the MPU, which is usually being used for memory protection in
safety-critical embedded systems, is impractical as it does not meet the fine-grained memory
protection requirements. Configuring the MPU takes a considerable number of cycles in kernel mode
and is inefficient when it comes to the power consumption and die area.

CHERIoT~\cite{amar2023cheriot} is targeting low-end IoT devices and aims to secure them using CHERI. It uses
similar GP-relative addressing to CompartOS, however, CHERIoT is not designed for dynamic systems or high-end
embedded systems such as embedded Linux.

Huawei's CheriLinux~\cite{huaweicherilinux}, though completely separate, is the closest to our work, specifically only to our first contribution (CHERI C Linux). They target MMU-based server-class Linux, unlike our
work here which targets MMU-less Linux. Furthermore, they only support CHERI C systems without software compartmentalisation at all, which we provide as three separate contributions.

\section{Conclusion}
We have shown how MMU-less Linux, one of the most advanced high-end operating systems, could be secured using two CHERI protection models: CHERI C and compartmentalisation. CHERI C works at the language
level by protecting every pointer in the kernel and userspace. This provides complete spatial pointer safety. Compartmentalisation splits up a monolithic system such as the kernel and userspace into smaller and isolated logical compartments
such as kernel modules in the kernel, and user applications and libraries in userspace. This reduces the effects of future unknown attacks.

We have demonstrated the effort it took in order to secure the upstream RISC-V MMU-less Linux, \textit{Busybox} and \textit{uclibc-ng} sysetms. The \gls{LoC} changes were quite minimal for such large code bases written in millions LoC, which emphasises the
practicality and compatibility of securing low-level software using CHERI with minimum effort. This completely relies on CHERI as the sole hardware protection mechanism. We further show that the CompartOS model~\cite{almatary2022compartos, UCAM-CL-TR-976} is applicable to large high-end embedded systems such as Linux.

\section{Acknowledgements}
This work was supported by the Defense Advanced Research
Projects Agency (DARPA) under contract HR001123C0031
(“MTSS”). The views, opinions, and/or findings contained
in this report are those of the authors and should not be
interpreted as representing the official views or policies
of the Department of Defense or the U.S. Government.
Distribution Statement A. Approved for public release: distribution is unlimited.

\bibliography{cherielinux}

\end{document}